\documentclass[11pt, a4paper, logo, copyright]{mint}

\pdftrailerid{redacted}

\usepackage[authoryear, sort&compress, round]{natbib}

\usepackage{times}
\usepackage{latexsym}
\usepackage{array}
\usepackage{algorithm}
\usepackage{algpseudocode}
\usepackage{amsmath}
\usepackage{wrapfig}

\usepackage[T1]{fontenc}

\usepackage[utf8]{inputenc}

\usepackage{microtype}

\usepackage{fix-cm}  

\usepackage{graphicx}
\usepackage[most]{tcolorbox}
\usepackage{url}
\usepackage{amsmath, amssymb}
\usepackage{multirow} 
\usepackage[table]{xcolor}
\usepackage{booktabs}
\usepackage{pifont}

\usepackage{enumitem}
\usepackage{ulem}
\usepackage{amsmath,amssymb,amsthm}
\usepackage{pifont}
\usepackage{graphicx}
\usepackage{pifont}
\usepackage{subcaption}
\usepackage{url}
\usepackage{xcolor}
\usepackage[table]{xcolor} 

\usepackage{graphicx}
\usepackage{hyperref}
\usepackage{tabularx}
\usepackage{forest}
\usepackage{booktabs}
\usepackage{amssymb}
\usepackage{makecell}
\usepackage{pifont}

\usepackage[utf8]{inputenc} 
\usepackage[T1]{fontenc}    

\definecolor{emory}{HTML}{002878}

\usepackage[colorlinks=true]{hyperref}
\hypersetup{
    linkcolor=emory,       
    citecolor=emory,      
    filecolor=emory,    
    urlcolor=emory         
}
\usepackage{url}            
\usepackage{amsfonts}       
\usepackage{nicefrac}       
\usepackage{microtype}      
\usepackage{colortbl}         
\usepackage{enumitem}
\usepackage[nameinlink,capitalize,noabbrev]{cleveref}
\usepackage{titletoc}
\usepackage{lipsum}

\definecolor{lightlav}{HTML}{EDE7F6}
\definecolor{darklav}{HTML}{5E35B1}
\usepackage{graphicx}
\usepackage{siunitx}
\usepackage{enumitem}
\usepackage{pifont}
\usepackage{amsthm}

\theoremstyle{definition}

\usepackage{wrapfig}
\usepackage{caption}
\usepackage[table]{xcolor}
\definecolor{headercolor}{RGB}{219,217,226}

\usepackage{hyperref}      
\usepackage{fontawesome5}  
\usepackage{academicons}   
\usepackage{xcolor}        

\usepackage{booktabs}
\usepackage{multirow}
\usepackage{threeparttable}
\usepackage[table]{xcolor}
\usepackage{siunitx}
\usepackage{makecell}
\definecolor{lightgreen}{RGB}{226,239,218}
\definecolor{lightred}{RGB}{252,228,214}
\definecolor{lightgray}{RGB}{245,245,245}
\definecolor{llmgreen}{RGB}{32,110,70}
\definecolor{llmred}{RGB}{180,60,50}
\newcommand{\best}[1]{\textbf{#1}}
\newcommand{\gain}[1]{\textcolor{llmgreen}{\scriptsize(+#1)}}

\newcommand{\costdown}[1]{\textcolor{llmred}{\scriptsize($\downarrow$#1)}}

\title{Scaling Teams or Scaling Time? \\Memory Enabled Lifelong Learning in LLM Multi-Agent Systems}

\newcommand*{\affmark}[1][*]{\textsuperscript{#1}}

\author[1]{Shanglin Wu\affmark[$\dagger$]}
\author[1]{Yuyang Luo}
\author[2]{Yueqing Liang}
\author[3]{Kaiwen Shi}
\author[3]{Yanfang Ye}
\author[4]{Ali Payani}
\author[1]{Kai Shu\affmark[*]}

\affil[1]{Emory University}
\affil[2]{Illinois Institute of Technology}
\affil[3]{University of Notre Dame}
\affil[4]{Cisco Research}

\affil[ ]{\par\affmark[$\dagger$] First Author\\
\affmark[*] Corresponding Author}

\begin{abstract}
Large language model (LLM) multi-agent systems can scale along two distinct dimensions: by increasing the number of agents and by improving through accumulated experience over time. Although prior work has studied these dimensions separately, their interaction under realistic cost constraints remains unclear. In this paper, we introduce a conceptual scaling view of multi-agent systems that jointly considers team size and lifelong learning ability, and we study how memory design shares this landscape. To this end, we propose \textbf{LLMA-Mem}, a lifelong memory framework for LLM multi-agent systems under flexible memory topologies. We evaluate LLMA-Mem on \textsc{MultiAgentBench} across coding, research, and database environments. Empirically, LLMA-Mem consistently improves long-horizon performance over baselines while reducing cost. Our analysis further reveals a non-monotonic scaling landscape: larger teams do not always produce better long-term performance, and smaller teams can outperform larger ones when memory better supports the reuse of experience. These findings position memory design as a practical path for scaling multi-agent systems more effectively and more efficiently over time. Our code is available at \url{https://github.com/ShanglinWu/MAS_lifelong_learning}.

\end{abstract}

\begin{document}

\maketitle

\begin{figure}[htbp!]
\centering
\includegraphics[width=1.0\textwidth]{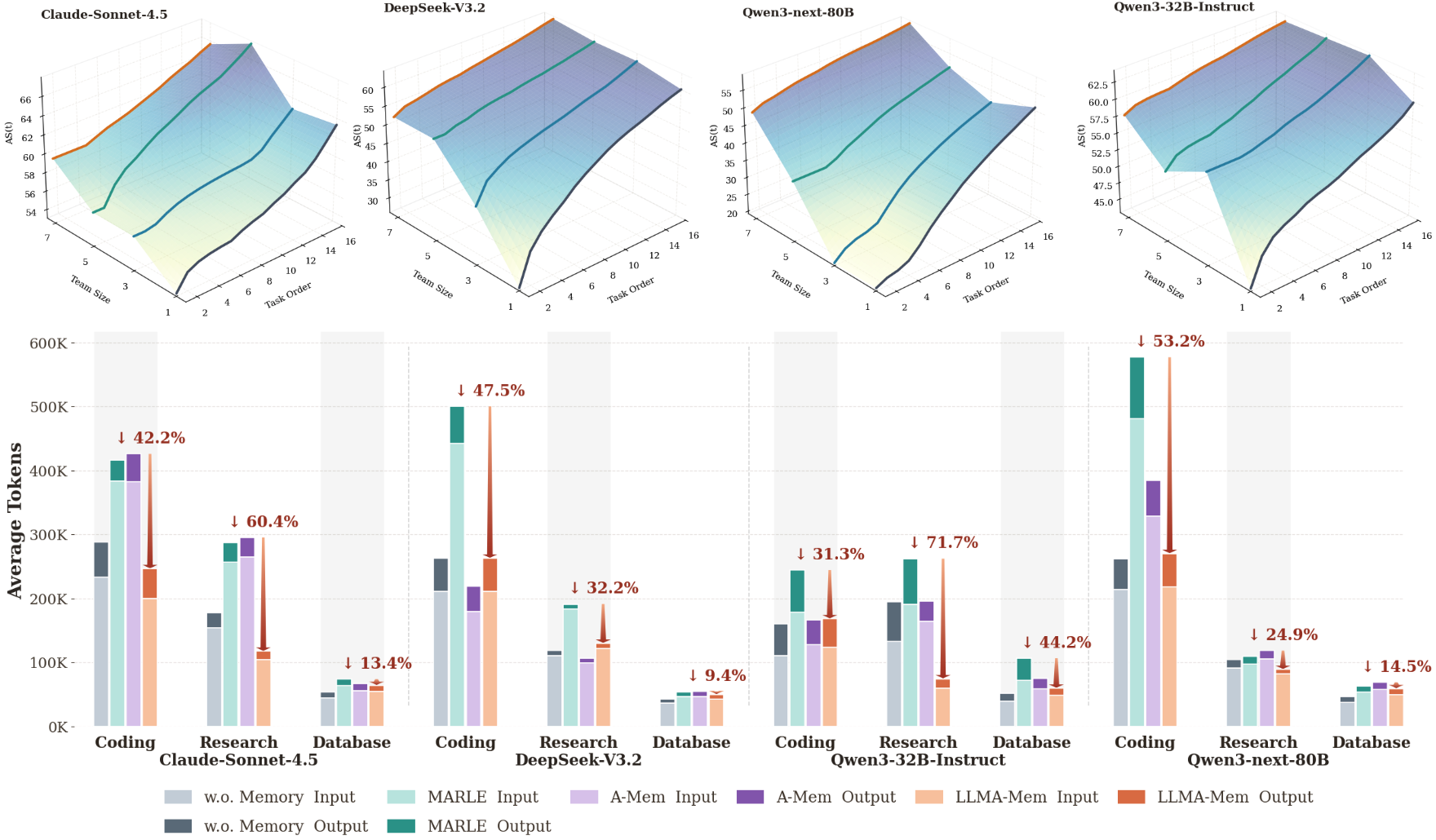}
\caption{Scaling space and cost comparison of LLMA-Mem. Top: LLMA-Mem enabled scaling space for multi-agent systems. Bottom: Average token usage per task, where LLMA-Mem shows substantial cost reduction compared to baselines across models.}
\label{fig: scaling space and cost comparison}
\end{figure}

\newpage
\section{Introduction}
Large language model (LLM)-based agent systems have demonstrated remarkable capabilities across diverse domains, from complex reasoning and code generation to embodied task execution and scientific discovery \citep{chen2024agentverse, wu2024autogen}. These systems extend beyond passive language generation to exhibit autonomous decision-making, tool use, and environmental interaction. A natural extension of this paradigm involves scaling from individual agents to multi-agent systems (MAS), where multiple LLM agents collaborate to tackle problems that exceed the capacity of a single agent. Such architectures prove essential for tasks requiring diverse expertise, extended context processing, parallel execution, or collective verification to reduce hallucinations \citep{chen2024more, du2023improving}. Beyond scaling team size, a distinct dimension of scaling emerges in lifelong learning \citep{zheng2025lifelong}, where agents continuously adapt and improve over time.

Current research predominantly addresses these two scaling dimensions in isolation. Studies on team-size scaling investigate how performance varies with the number of agents, revealing that collaborative scaling can follow logistic patterns \citep{qian2024scaling} and that smaller models with more agents can sometimes outperform larger models with fewer agents \citep{li2024more}. Separately, research on lifelong learning mainly focuses on how individual agents improve through accumulated experience, often through increasingly sophisticated memory architectures \citep{zheng2025lifelong}. However, the intersection of these dimensions remains largely unexplored. In practice, once a system moves from a single agent to a multi-agent team, lifelong learning is shaped not only by the quality of memory, but also by coordination overhead, communication cost, and information fragmentation \citep{kim2025towards}. This observation motivates the central perspective of our paper. We argue that LLM multi-agent systems should be viewed through a conceptual \textbf{scaling space} jointly defined by team size and lifelong learning ability. In this view, scaling is no longer a single-axis question of whether ``more agents are better'' or ``more experience is helpful''. Instead, it becomes a systems question about how the interaction between team composition and learning over time shapes long-horizon effectiveness and efficiency. Our experiments make this tension explicit: larger teams do not always achieve better long-term performance, and better memory can change the outcome enough that a smaller team becomes preferable. This reveals a new trade-off between scaling horizontally through more agents and scaling temporally through better learning over time.

Among the many potential factors, we focus on memory as the primary mechanism governing lifelong learning ability, because memory determines how agents retain, retrieve, and build upon past experiences \citep{zhang2025survey, zheng2025lifelong}. The difficulty of memory management grows with memory complexity, which depends on both team size and communication mechanism \citep{kim2025towards}. We further argue that memory topology, which determines who can read or write which memories, is also a crucial factor in this complexity \citep{wu2025memory}. Even under identical communication structures, different topologies distribute complexity very differently across the system. Existing work mainly studies isolated memory designs and does not sufficiently address cross-task learning in multi-agent settings. This leads to our first research question: \textbf{RQ1: How can memory design improve lifelong learning in multi-agent systems?} To answer it, we propose \textbf{LLMA-Mem}, a multi-agent memory framework that distills compact procedural memories from episodic experience while also modeling team capabilities through transactive memory. This design aims to improve both long-horizon performance and efficiency: in our experiments, LLMA-Mem improves lifelong learning ability over the baseline in all model-environment settings while reducing token usage by \costdown{9.4\%} to \costdown{71.7\%} relative to competing memory baselines. To study these effects, we require benchmarks that go beyond static question answering or single-turn reasoning, with a focus on multi-agent scenarios. We therefore select \textsc{MultiAgentBench} \citep{zhu2025multiagentbench}, whose multi-step environments expose both task completion and inter-agent communication over sequential tasks.

Following the previous discussion, our second research question is \textbf{RQ2: How does lifelong learning ability interact with team size in multi-agent systems?} To answer \textbf{RQ2}, we study LLMA-Mem over long task sequences and explicitly quantify how increasing team size changes the marginal value of experience. Intuitively, larger teams can provide stronger parallelism, which may accelerate learning early on. At the same time, scaling the team introduces coordination overhead, duplicated effort, and fragmented information flow, all of which can weaken long-horizon gains. Our results confirm that this relationship is not monotonic: smaller teams can outperform larger ones when memory supports stronger accumulation and reuse of experience. Through systematic analysis, we demonstrate how the interplay between memory design and team size creates a nuanced landscape for agentic scaling, offering guidance for researchers and practitioners in designing effective multi-agent systems. Our main contributions are as follows:
\begin{itemize}
\item We introduce a joint scaling perspective for LLM multi-agent systems that connects team size scaling and lifelong learning scaling, framing them as an interacting scaling space rather than two isolated axes.
\item We propose \textbf{LLMA-Mem}, a lifelong memory framework that separates episodic memory as experiential substrate, procedural memory as the main mechanism for cross-task transfer, and transactive memory as the mechanism for team-level capability and coordination modeling.
\item We conduct systematic empirical study of how lifelong learning interacts with team size in LLM multi-agent systems, revealing a non-monotonic scaling pattern.
\item We show that LLMA-Mem improves long-horizon performance while reducing token cost relative to existing memory baselines, highlighting memory design as a practical path toward more effective and more efficient multi-agent systems.
\end{itemize}

\section{Related Works}
\subsection{Scaling of team size in LLM Multi-agent Systems}
A central question in designing LLM-based multi-agent systems (MAS) is how many agents to deploy in the systems, yet recent evidence suggests that team size scaling is highly conditional rather than universally monotonic. Early work on scalable collaboration networks argues that increasing the number of collaborating agents can produce systematic gains that follow a saturating scaling pattern. In particular, MacNet \citep{qian2024scaling} reports a logistic shaped collaborative scaling law with diminishing returns beyond moderate team sizes, and further highlights that topology and message policy can matter as much as raw agent count, with irregular connectivity sometimes outperforming regular structures. Complementing this optimistic view, \citet{kim2025towards} provide a large controlled study across multiple agentic benchmarks and canonical coordination architectures, showing that scaling team size can be net negative under fixed budgets due to coordination overhead, tool call contention, and error amplification, especially in long-horizon tasks. This raises concerns that some reported multi-agent gains may be confounded by weak baselines. \citet{xu2026rethinking} argue that when agents are homogeneous, a single agent can often simulate the workflow through multi-turn execution while benefiting from KV-cache reuse, substantially narrowing or eliminating the apparent advantage of larger teams. Finally, MAS-Orchestra \citep{ke2026mas} reframes team size as an orchestration variable rather than a fixed hyperparameter, introducing controlled benchmarks and a notion of degree of MAS to study when richer multi-agent structures are necessary. Despite this progress, the literature still lacks a unified understanding of how team size should be chosen for systems that evolve over time. Existing scaling studies largely measure one-off task performance or short horizon coordination, leaving open how team size interacts with information fragmentation, and the accumulation of reusable knowledge across task sequences.

\subsection{Lifelong learning in LLM Agent Systems}
Beyond single-task performance, recent work studies lifelong learning in LLM agents, aiming for continual improvement without repeated retraining. \citet{zheng2025lifelong} conceptualize lifelong agents through the interaction of perception, memory, and action, emphasizing the transformation of experience into transferable competence. In practice, many systems implement lifelong adaptation through external memory mechanisms. For example, Agent Workflow Memory (AWM) induces reusable workflows from past trajectories to guide future decisions \citep{wang2024agent}, while ReasoningBank extracts generalizable reasoning memories from both successes and failures and leverages memory aware test-time scaling to enrich experience in single agent scenario \citep{ouyang2025reasoningbank}. As memory accumulates, efficiency becomes a challenge. Systems such as SimpleMem introduce structured compression and intent aware retrieval to support long term interaction under bounded token budgets \citep{liu2026simplemem}. While works like MemSkill \citep{zhang2026memskill} and Evo-Memory \citep{wei2025evo} models memory writing, pruning, and retrieval as evolving capabilities. Despite these advances, most lifelong agent research remains single agent centric, leaving open how lifelong learning behaves as team size grows and how memory topology influences the trade-off between scaling agents and scaling time.

\subsection{Memory in LLM-based Agent Systems}
Memory mechanisms have become foundational to LLM-based agents, enabling them to transcend the stateless nature of base language models and develop persistent, adaptive behaviors. Early influential work by \cite{park2023generative} introduced Generative Agents, which maintain a memory stream of experiences, synthesize higher level reflections, and dynamically retrieve relevant memories for planning. MemGPT \citep{packer2023memgpt} proposed a hierarchical two-tier architecture inspired by operating system virtual memory, with a main context analogous to RAM and external storage analogous to disk, enabling agents to self edit their memory through function calls. More recent work like A-MEM \citep{xu2025mem} introduces dynamic memory organization with interconnected knowledge networks. The extension of memory to multi-agent systems introduces unique challenges distinct from single agent memory \citep{wu2025memory}. Transactive memory systems become essential for efficient task allocation \citep{zhang2025survey}. Memory topology emerges as a critical design dimension, with three primary configurations identified in the literature: per-agent local memory where each agent maintains private stores, centralized shared memory using blackboard-style architectures, and hybrid approaches combining local perceptual memory with shared summarized world-state \citep{wu2025memory}. Collaborative Memory frameworks \citep{rajaram2010collaborative} formalize this with two-tier architectures featuring private fragments and shared fragments governed by dynamic access controls. LEGOMem  \citep{han2025legomem} specifically addresses procedural memory in multi-agent workflows, with orchestrator memory for high level planning and fine-grained subtask retrieval for worker agents. Despite these advances, current multi-agent memory research primarily focuses on coordination efficiency within individual tasks rather than cross-task evolution, leaving substantial gaps in understanding how memory architectures enable teams to improve over successive experiences.

\section{Method}
\label{sec: method}

\subsection{Preliminary Study}
\begin{wrapfigure}{r}{0.60\textwidth} 
  \centering
  \includegraphics[width=0.60\textwidth]{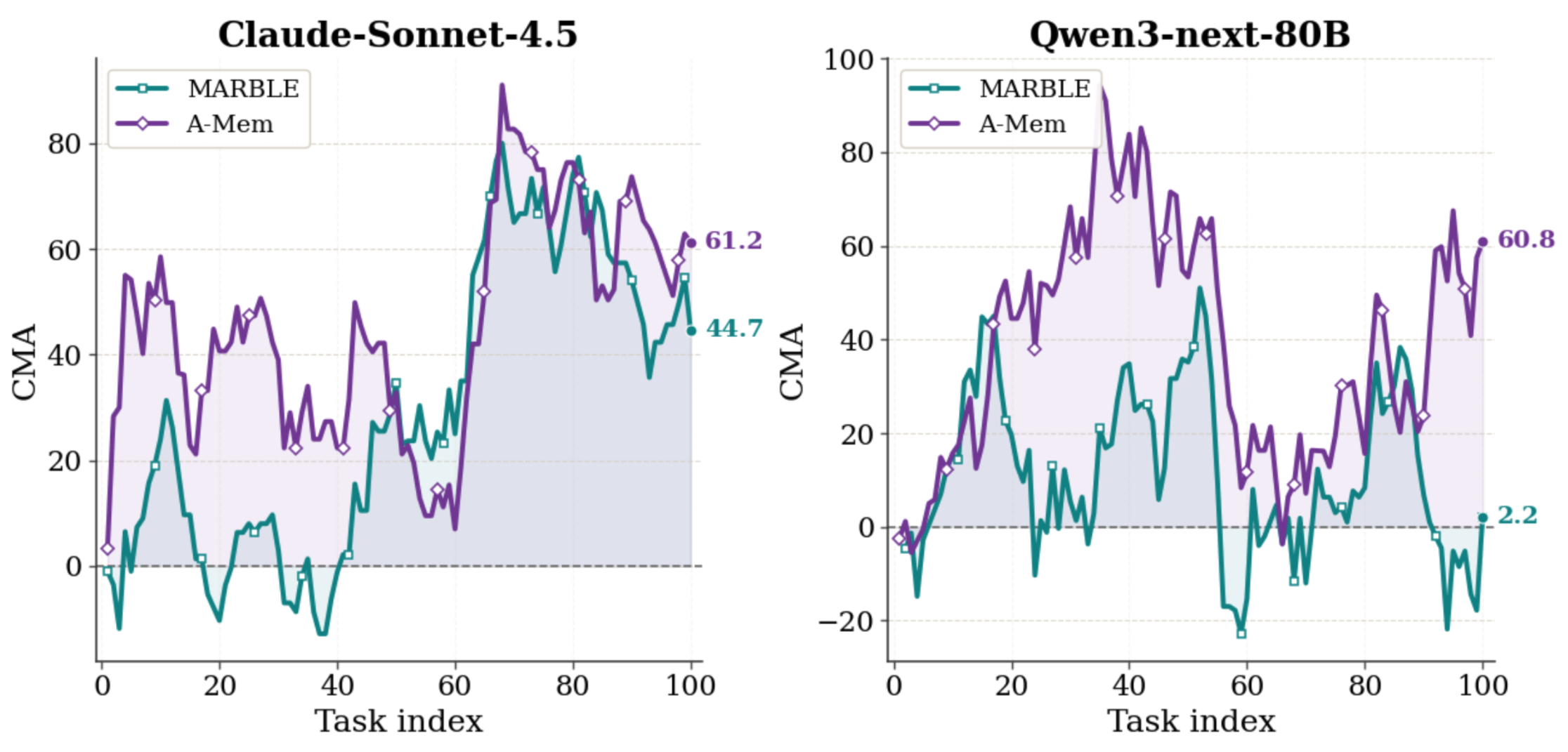}
  \caption{CMA curves for Claude-Sonnet-4.5 and Qwen3-next-80B, showing limited long-horizon adaptation.}
  \label{fig:preliminary}
\end{wrapfigure}

Before presenting LLMA-Mem, we first conduct a preliminary study on two existing frameworks, MARBLE \citep{zhu2025multiagentbench} and A-Mem \citep{xu2025mem}, to examine whether existing designs are sufficient for lifelong learning in multi-agent systems. We choose these two baselines because they reflect two different design patterns. MARBLE adopts a more general memory design for MAS, combining short-term memory, long-term memory, and shared memory to support task execution and inter-agent interaction. In contrast, A-Mem is more memory-evolution oriented, where a dedicated memory agent explicitly manages, organizes, and refines memories over time. To evaluate their lifelong learning behavior, we analyze the cumulative performance gain (CMA) on coding tasks which is defined in Section \ref{sec: experiments}, using the no-memory setting as the reference baseline. Figure~\ref{fig:preliminary} shows the CMA curves for Claude-Sonnet-4.5 and Qwen3-next-80B. If an existing memory design is sufficient for lifelong MAS, we would expect its cumulative gain to grow steadily as the system accumulates more task experience. However, the observed trends suggest that the gains of MARBLE and A-Mem are still limited. Although both methods can outperform the no-memory baseline in parts of the task sequence, the improvement is not consistently amplified over time, and the cumulative gain often grows slowly or plateaus.

This suggests two key limitations of existing methods. \textbf{First, existing methods do not sufficiently support cross-task reuse.} The empirical trends imply that stored experiences are not consistently converted into compact and reusable knowledge that can benefit future tasks. In MARBLE, memory mainly serves as a general storage mechanism for task-related information, while A-Mem emphasizes memory refinement and dynamic organization. However, neither design explicitly separates raw task experiences from higher-level reusable procedures in a way that is tailored for continual multi-task transfer. For lifelong MAS, memory should not only preserve past information, but also progressively distill repeated successful experiences into concise procedural knowledge. Moreover, in multi-agent settings, memory should include team-state awareness, so that agents can reason from who is capable of what and how collaborators have performed in previous interactions. \textbf{Second, existing methods do not explicitly address how memory topology affects lifelong learning complexity.} Both MARBLE and A-Mem are instantiated under a fixed memory topology, while in lifelong MAS the choice of topology directly affects memory scalability, sharing efficiency, and role specialization. A local topology preserves agent-specific knowledge and specialization, but introduces fragmented memory growth as team size increases. A shared topology promotes collective reuse, but may increase retrieval noise and memory management overhead as more agents write into the same store. A hybrid topology may potentially balance these trade-offs by preserving local experiences while sharing abstracted knowledge and coordination statistics. The challenge lies in both storing and distributing memory across agents as the system evolves.

\begin{figure}[htbp!]
\centering
\includegraphics[width=1.0\textwidth]{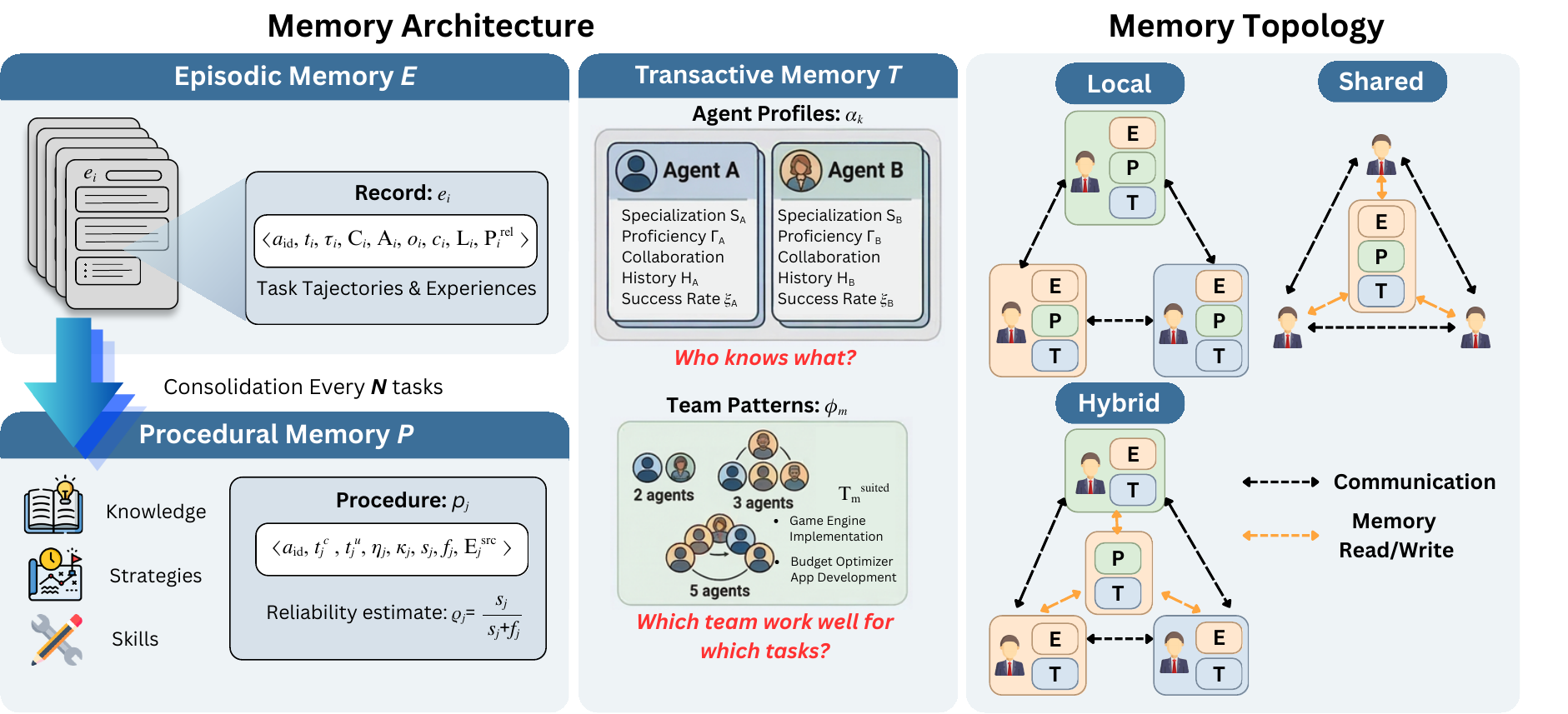}
\caption{LLMA-Mem maintains three memory components: episodic memory for task experiences, procedural memory for consolidated reusable strategies, and transactive memory for agent capabilities and team coordination. The right panel illustrates three memory topology configurations that determine how memory is distributed and accessed across agents.}
\label{fig: architecture & topology}
\end{figure}

\subsection{Memory Architecture}
Based on this preliminary study, we propose \textbf{LLMA-Mem} (\textbf{L}ife\textbf{L}ong \textbf{M}ulti-\textbf{A}gent \textbf{Mem}ory), a unified memory framework designed to support lifelong learning across diverse multi-agent topologies. The framework consists of three interdependent memory modules: episodic, procedural, and transactive. Each serving distinct but complementary roles in enabling lifelong learning. To ensure reproducibility, we present detailed settings and prompts in Appendix \ref{app:llmamem_design} and Appendix \ref{app:prompts}. 

As illustrated in figure \ref{fig: architecture & topology}, LLMA-Mem maintains three complementary memory components: \emph{episodic memory}, \emph{procedural memory}, and \emph{transactive memory}. Episodic memory records task-level experiences in their original context, procedural memory abstracts reusable strategies from accumulated experience, and transactive memory models agent capabilities and team coordination patterns. Together, these components enable both individual adaptation and system-level lifelong learning in multi-agent environments. We show a detailed case study on coding task at Appendix \ref{app:llmamem_design}.

\textbf{Episodic Memory.} Episodic memory stores rich records of past task executions, including the task context, agent interactions, outcomes, and post hoc reflections. It serves as the foundation for later abstraction and supports case-based reasoning when the system encounters situations similar to prior experiences. Formally, an episode $e_i$ is represented as
$e_i = \langle a_{\mathrm{id}}, t_i, \tau_i, \mathcal{C}_i, \mathcal{A}_i, o_i, c_i, \mathcal{L}_i, \mathcal{P}^{\mathrm{rel}}_i \rangle$,
where $a_{\mathrm{id}}$ denotes the agent ID, $t_i$ is the timestamp, $\tau_i$ is the task description, $\mathcal{C}_i$ is the team composition, $\mathcal{A}_i$ is the action sequence, $o_i$ is the task outcome, $c_i$ denotes the environmental context, $\mathcal{L}_i$ contains extracted lessons, and $\mathcal{P}^{\mathrm{rel}}_i$ links the episode to related procedures. By preserving full task trajectories rather than only final outcomes, episodic memory provides the raw experiential substrate from which higher-level knowledge can later be consolidated.

\textbf{Procedural Memory.} Procedural memory stores reusable strategies and skills abstracted from episodic experiences. In contrast to episodic memory, which is instance-specific, procedural memory captures generalized knowledge that has proven effective across multiple tasks. This abstraction improves retrieval efficiency and facilitates transfer to future tasks. A procedure $p_j$ is defined as
$p_j = \langle a_{\mathrm{id}}, t^{c}_j, t^{u}_j, \eta_j, \kappa_j, s_j, f_j, \mathcal{E}^{\mathrm{src}}_j \rangle$,
where $a_{\mathrm{id}}$ is the owner or creator, $t^{c}_j$ and $t^{u}_j$ are the creation and most recent update times, $\eta_j$ is the procedure title, $\kappa_j$ is the procedural knowledge content, $s_j$ and $f_j$ are the numbers of successful and failed applications, and $\mathcal{E}^{\mathrm{src}}_j$ denotes the supporting source episodes. We estimate the reliability of procedure $p_j$ by its empirical success rate: $\rho_j = \frac{s_j}{s_j + f_j}$. In practice, procedural memory is updated through consolidation over episodic memory, transforming repeated successful patterns into explicit action templates.

\textbf{Transactive Memory.} Transactive memory captures "who knows what" within the multi-agent system and "which team configurations work well for which tasks". This component is particularly important in multi-agent lifelong learning, where effective task allocation depends not only on individual competence but also on accumulated collaboration experience. We model transactive memory using two structures: agent profiles and team patterns. An agent profile for agent $k$ is defined as
$\alpha_k = \langle a_{\mathrm{id}}, \mathcal{S}_k, \Gamma_k, \mathcal{H}_k, \xi_k \rangle$, where $\mathcal{S}_k$ denotes the agent's specialization areas, $\Gamma_k$ maps task types to proficiency estimates, $\mathcal{H}_k$ records collaboration history with other agents, and $\xi_k$ is an overall success rate. A team pattern for team configuration $m$ is defined as $\phi_m = \langle \mathcal{C}_m, \mathcal{T}^{\mathrm{suited}}_m \rangle$, where $\mathcal{C}_m$ is the team composition and $\mathcal{T}^{\mathrm{suited}}_m$ is the set of task types for which this team is effective. During planning, the system consults transactive memory to support task decomposition, role assignment, and team formation. This component is therefore central to our analysis of how team size affects lifelong learning and coordination efficiency.

\subsection{Memory Topology Configurations}

As shown in figure \ref{fig: architecture & topology}, LLMA-Mem supports three memory topologies that determine how knowledge is distributed and shared across agents. \textbf{Local Topology:} In the local topology, each agent $a_i$ maintains a private memory store $\mathcal{M}_i = \langle \mathcal{E}_i, \mathcal{P}_i, \mathcal{T}_i \rangle$, where $\mathcal{E}_i$, $\mathcal{P}_i$, and $\mathcal{T}_i$ denote the agent's episodic, procedural, and transactive memory, respectively. Agents cannot directly access one another's memory, and knowledge transfer occurs only through explicit inter-agent communication during task execution. \textbf{Shared Topology:} In the shared topology, all agents access a centralized memory store $\mathcal{M}_{\mathrm{shared}} = \langle \mathcal{E}_{\mathrm{shared}}, \mathcal{P}_{\mathrm{shared}}, \mathcal{T}_{\mathrm{shared}} \rangle.$ All newly acquired experiences contribute to a common memory pool, allowing the entire team to access collective experience. \textbf{Hybrid Topology:} The hybrid topology combines private and shared memory: $\mathcal{M}^{\mathrm{hybrid}}_i =
\langle
\mathcal{E}^{\mathrm{local}}_i,\,
\mathcal{T}^{\mathrm{local}}_i,\,
\mathcal{P}_{\mathrm{shared}},\,
\mathcal{T}_{\mathrm{shared}}
\rangle.$ Here, episodic memory remains local to preserve agent-specific experience, while procedural memory and aggregated team statistics are shared across agents. Among three topologies, local topology enables role-specific memory storage, while shared topology enables cross-agent memory reuse. Hybrid topology balances knowledge sharing and specialization where agents retain their own task histories, while the team collectively benefits from abstracted procedures and global coordination statistics.

\begin{figure}[htbp!]
\centering
\includegraphics[width=1.0\textwidth]{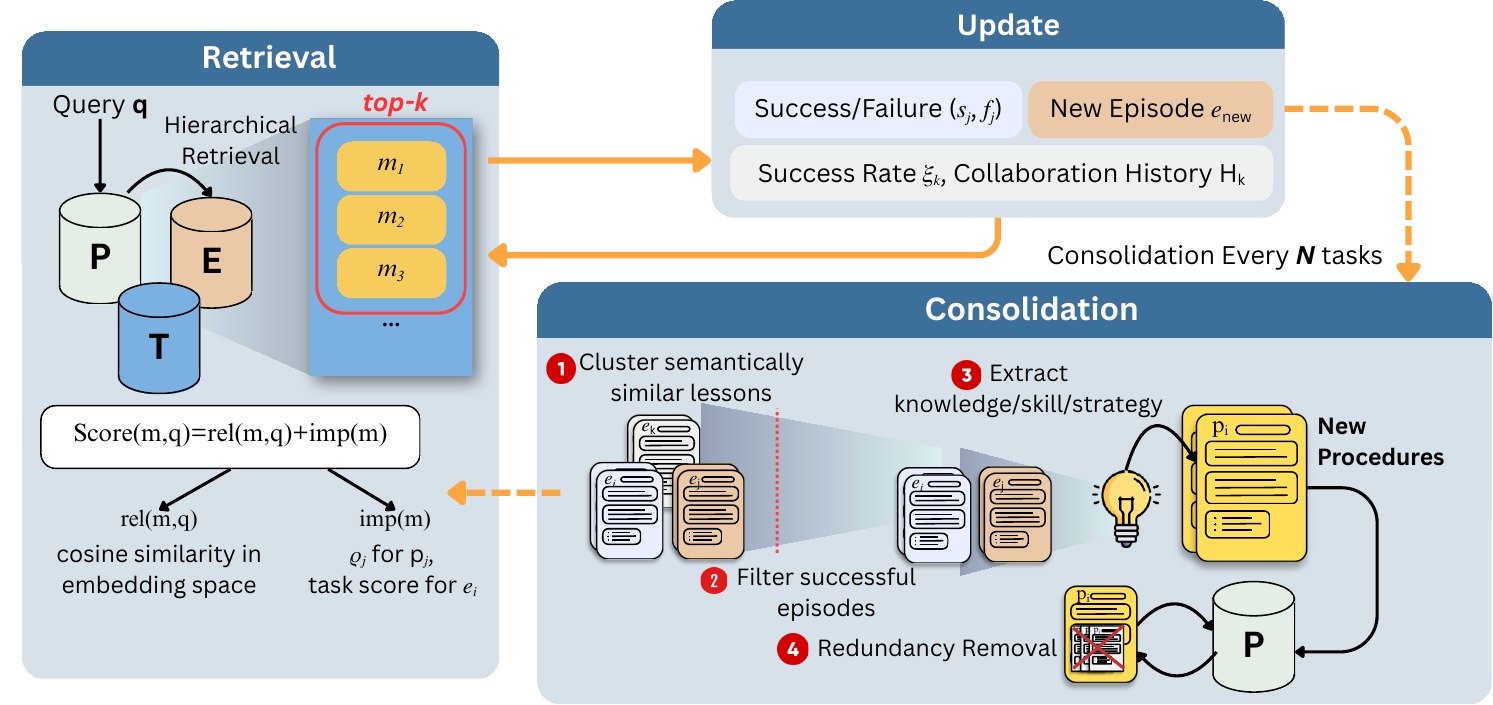}
\caption{During task execution, the system retrieves relevant memories using a relevance–importance score, updates episodic and transactive statistics after each task, and periodically consolidates episodic experiences into reusable procedural knowledge.}
\label{fig: lifecycle}
\end{figure}

\subsection{Memory Lifecycle}
As shown in figure \ref{fig: lifecycle}, the memory lifecycle in LLMA-Mem consists of three phases: retrieval, update, and consolidation. During task execution, the system retrieves memory entries relevant to the current query context $q$ (e.g., the task description). Retrieval follows a hierarchical strategy: the agent first queries procedural memory when available, and falls back to episodic memory if no sufficiently relevant procedure is found. For a memory item $m$, the retrieval score is defined as
\begin{equation}
\mathrm{score}(m,q)
=
\mathrm{rel}(m,q)
+
\mathrm{imp}(m)
\end{equation}
where $\mathrm{rel}(m,q)$ denotes semantic relevance and $\mathrm{imp}(m)$ denotes memory importance. Semantic relevance is computed via cosine similarity in embedding space. For procedural memory importance is defined as success rate $\rho_j$, and for episodic memory importance is defined as task score. Both semantic relevance and importance are standardized.

In our implementation, the top-$k$ retrieved items are provided to the agent as context. 

After each task, the memory system is updated at three levels. First, a new episode $e_{\mathrm{new}}$ is appended to episodic memory:
\begin{equation}
\mathcal{E}^{(t+1)} = \mathcal{E}^{(t)} \cup \{e_{\mathrm{new}}\}.
\end{equation}
Second, for each procedure used during the task, success and failure counts are incrementally updated:
\begin{equation}
s_j^{(t+1)} = s_j^{(t)} + \mathbb{1}_{\mathrm{success}},
\qquad
f_j^{(t+1)} = f_j^{(t)} + \mathbb{1}_{\mathrm{failure}}.
\end{equation}
These statistics are then used to refresh the reliability estimate of the procedure. Third, transactive memory is updated using the observed task outcomes. For example, the reliability of agent $k$ is updated as
\begin{equation}
\xi_k^{(t+1)} =
\frac{\mathrm{successes}_k}{\mathrm{total\_tasks}_k},
\end{equation}
while the agent's specialization estimates and the team-level statistics are updated according to the latest execution trace. To convert raw experiences into reusable knowledge, LLMA-Mem periodically consolidates episodic memory into procedural memory every $N$ episodes. The system groups semantically similar episodes according to their extracted lessons, identifies recurring successful patterns, and abstracts them into candidate procedures, as summarized in Algorithm~\ref{alg:procedure_extraction}. This consolidation stage is the key mechanism by which LLMA-Mem transforms task-level experience into reusable and shareable knowledge, enabling continual improvement over time rather than treating tasks independently. 

\begin{algorithm}[h]
\caption{Procedural Knowledge Extraction}
\label{alg:procedure_extraction}
\begin{algorithmic}[1]
\State \textbf{Input:} episodic memory $\mathcal{E}$
\State \textbf{Output:} new procedures $\mathcal{P}_{\mathrm{new}}$
\State Cluster episodes in $\mathcal{E}$ by semantic similarity of \textit{lessons\_learned}.
\For{each cluster $C$ with $|C| \geq 2$}
    \State $E_{\mathrm{succ}} \gets \{e \in C \mid e \text{ is successful}\}$
    \If{$|E_{\mathrm{succ}}| \geq 2$}
        \State Extract a generalized strategy $\kappa$ from $E_{\mathrm{succ}}$
        \State Create procedure $p_{\mathrm{new}}$ with source set $\mathcal{E}^{\mathrm{src}} \gets E_{\mathrm{succ}}$
        \State Add $p_{\mathrm{new}}$ to $\mathcal{P}_{\mathrm{new}}$
    \EndIf
\EndFor
\State Remove redundant procedures whose source episodes are subsets of others
\State \Return $\mathcal{P}_{\mathrm{new}}$
\end{algorithmic}
\end{algorithm}

\section{Experiments}
\label{sec: experiments}

\subsection{Experimental Setup}

\textbf{Benchmark.}
To evaluate the lifelong learning capability of LLMA-Mem, we adopt an agentic benchmark which better captures the interactive and sequential nature of real-world agent systems, making them particularly suitable for studying memory-enabled learning dynamics. We therefore choose \textsc{MultiAgentBench} \citep{zhu2025multiagentbench}, a benchmark specifically designed for evaluating LLM multi-agent systems across diverse environments. \textsc{MultiAgentBench} contains both collaborative and competitive settings. Since our memory framework is designed to support agent teams that cooperate to achieve shared goals, we focus on the collaborative environments. In particular, we select three representative domains: \textit{coding}, \textit{research}, and \textit{database} with 100 tasks for each environment.

\noindent\textbf{Models.}
To study the robustness of the proposed memory framework across different model families and parameter scales, we evaluate both closed-source and open-source LLMs. Specifically, we use Claude-Sonnet-4.5 \citep{anthropic2025claude}, DeepSeek-V3.2 \citep{liu2025deepseek}, Qwen3-next-80B \citep{qwen2025qwen3next}, and Qwen3-32B-Instruct \citep{yang2025qwen3}. Within the \textsc{MultiAgentBench} framework, Claude-Sonnet-4.5 is used as the evaluation model to score task outcomes. For LLMA-Mem and A-Mem, we utilize Titan-text-embeddings-v2 \citep{aws2024titan_v2} as embedding model. All models are accessed through the Amazon Bedrock API under identical inference settings. To ensure reproducibility, we provide detailed experimental configurations in the Appendix \ref{app:baseline_details}.

\noindent \textbf{Baselines.} We choose three baselines compare to LLMA-Mem. (i) W.o. Memory: only utilize LLM into \textsc{MultiAgentBench} without using any memory module. (ii) MARBLE \citep{zhu2025multiagentbench}: the multi-agent collaboration framework proposed in \textsc{MultiAgentBench} with a memory module. MARBLE's memory module follows popular design protocol when both short-tern memory, long-term memory, and shared memory exist. (iii) A-Mem \citep{xu2025mem}: A-Mem is one of the representative memory framework which maintains a dynamic knowledge network and also implements a evolved memory storage, which is related and comparable to LLMA-Mem. For more baseline implementation details, see Appendix \ref{app:baseline_details}.

\noindent\textbf{Evaluation Metrics.} We evaluate lifelong multi-agent performance using the task-level scores provided by the original \textsc{MultiAgentBench} codebase \citep{zhu2025multiagentbench}. Specifically, \textsc{MultiAgentBench} reports two complementary metrics:
\textit{Task Score} (TS), which measures the correctness or quality of the final task output, and \textit{Communication Score} (CS), which evaluates inter-agent communication and planning during the task execution process. TS captures the agents' ability to solve the task itself, while CS reflects the quality of collaboration and coordination among agents. To obtain a unified task-level utility, we define
\begin{equation}
S_t = \frac{\mathrm{TS}_t + \mathrm{CS}_t}{2},
\end{equation}
\noindent where $S_t$ denotes the combined performance on task $t$. We evaluate lifelong learning ability through the evolution of performance over the task sequence. First, we analyze the performance trajectory $\{S_t\}_{t=1}^{T}$,
which reflects how task performance changes as the system accumulates experience over time. To obtain a smoother estimate of overall learning progress, we compute the running mean performance \citep{zheng2025lifelong, cai2021online}

\begin{equation}
AS(t) = \frac{1}{t}\sum_{i=1}^{t} S_i,
\end{equation}

\noindent which summarizes the average performance achieved up to task $t$.
We further report the Average of AS (AAS) as a scalar summary of
the system's performance across its entire lifetime \citep{caccia2022anytime, zheng2025lifelong}: 
\begin{equation}
AAS=\frac{1}{T} \sum_{i=1}^{T} AS(i)
\end{equation}

\noindent Finally, to quantify the benefit of memory, we compute the cumulative performance gain relative to the no-memory baseline \citep{zheng2025lifelongagentbench, serrano2021inter}:

\begin{equation}
CMA_t
=
\sum_{i=1}^{t}
\left(
S_i^{\mathrm{method}} - S_i^{\mathrm{nomem}}
\right).
\end{equation}

\noindent An increasing CMA curve indicates that accumulated memory continually consistently improves task performance over time. These metrics collectively capture both the instantaneous task performance and the long-term improvement dynamics of the multi-agent system.

\subsection{Main Results}

As shown in Table \ref{tab:main results}, LLMA-Mem is consistently competitive on overall task execution when we examine \textit{Task Score} (TS) and \textit{Communication Score} (CS) together. On TS, LLMA-Mem achieves the best or tied-best result in most model--environment pairs suche as Qwen3-next-80B on Research \gain{10.67}, indicating that consolidated memory usually transfers into stronger task completion quality. The trend on CS is more mixed: LLMA-Mem often improves collaboration, especially for DeepSeek-v3.2 on Research \gain{7.17}  and Qwen3-32B-Instruct on Database \gain{10.11}, but there are still cases such as Claude-Sonnet-4.5 on Research  and Qwen3-Next-80B on Research and Database where stronger task-solving does not fully translate into better communication quality. This suggests that memory helps agents act more effectively, while coordination remains more sensitive to model family and environment-specific interaction patterns.

\begin{table*}[htbp!]
      \centering
      \small
      \caption{Performance across environments. We report Task Score (TS), Coordination Score (CS), and Average of AS (AAS). Best results within each model and environment are boldfaced. For LLMA-Mem, the AAS cell also shows the change relative to the no-memory baseline.}
      \label{tab:main results}
      \resizebox{\textwidth}{!}{%
          \begin{tabular}{ll|ccc|ccc|ccc}
              \toprule
              \multirow{2}{*}{\textbf{Model}} & \multirow{2}{*}
  {\textbf{Setting}} &
              \multicolumn{3}{c|}{\textbf{Coding}} & \multicolumn{3}{c|}
  {\textbf{Research}} &
              \multicolumn{3}{c}{\textbf{Database}} \\
              & & TS & CS & AAS & TS & CS & AAS & TS & CS & AAS \\
              \midrule
              \multirow{4}{*}{Claude-Sonnet-4.5}
              & W.o. Memory & 64.43 & 42.29 & 53.42 & 75.12 & 78.97 & 76.94 &
  69.39 & 88.03 & 77.35 \\
              & MARBLE      & 63.05 & 44.71 & 53.99 & 75.17 & 81.84 & \best{78.37} &
  69.70 & 82.96 & 75.70 \\
              & A-Mem       & 61.57 & \best{46.41} & 53.76 & 75.69 & \best{82.47} & 77.79 &
  69.70 & 82.31 & 75.13 \\
              \rowcolor{lightgreen}
              & \textbf{LLMA-Mem}    & \best{65.00} & 44.75 & \best{54.02}~\gain{0.60} & \best{77.62} & 74.11 & 77.77~\gain{0.83} &
  \best{70.00} & \best{88.27} & \best{78.52}~\gain{1.17} \\
              \hline
              \multirow{4}{*}{Deepseek-v3.2}
              & W.o. Memory & 58.69 & 43.47 & 52.26 & 75.53 & 58.46 & 65.94 &
  71.00 & \best{82.40} & 75.55 \\
              & MARBLE      & 56.93 & 40.97 & 50.86 & \best{76.30} & 58.21 & 68.43 &
  70.00 & 75.60 & 72.62 \\
              & A-Mem       & 49.74 & 37.95 & 46.72 & 72.40 & 55.64 & 64.27 &
  68.69 & 73.48 & 70.51 \\
              \rowcolor{lightgreen}
              & \textbf{LLMA-Mem}    & \best{59.72} & \best{46.37} & \best{53.46}~\gain{1.20} & 75.10 & \best{65.63} & \best{71.86}~\gain{5.92} &
  \best{71.00} & 81.07 & \best{76.20}~\gain{0.65} \\
              \hline
              \multirow{4}{*}{Qwen3-next-80B}
              & W.o. Memory & 58.98 & 45.20 & 52.85 & 57.22 & 45.52 & 50.13 &
  69.00 & \best{72.32} & 67.61 \\
              & MARBLE      & 57.03 & 46.63 & 52.17 & 61.62 & 42.66 & 55.11 &
  67.00 & 68.53 & 64.38 \\
              & A-Mem       & 57.23 & \best{48.09} & 52.90 & \best{69.10} & \best{48.39} & \best{57.67} &
  \best{71.59} & 70.39 & \best{71.38} \\
              \rowcolor{lightgreen}
              & \textbf{LLMA-Mem}    & \best{60.65} & 44.86 & \best{53.12}~\gain{0.27} & 67.89 & 43.62 & 49.54 & 70.41
  & 68.55 & 69.98~\gain{2.37} \\
              \hline
              \multirow{4}{*}{Qwen3-32B-Instruct}
              & W.o. Memory & 49.24 & 44.23 & 48.01 & 71.56 & 72.63 & 72.06 &
  71.00 & 66.07 & 67.87 \\
              & MARBLE      & 48.82 & 44.40 & 46.76 & 72.38 & 68.62 & 70.36 &
  \best{71.72} & 69.15 & 69.68 \\
              & A-Mem       & 47.05 & \best{46.79} & 47.66 & \best{73.33} & 68.61 & 70.82 &
  71.43 & 73.23 & \best{71.20} \\
              \rowcolor{lightgreen}
              & \textbf{LLMA-Mem}    & \best{49.68} & 46.69 & \best{49.43}~\gain{1.42} & 73.17 & \best{73.08} & \best{73.50}~\gain{1.44} &
  70.41 & \best{76.18} & 71.06~\gain{3.19} \\
              \bottomrule
          \end{tabular}%
      }
  \end{table*}

\begin{figure}[htbp!]
\centering
\includegraphics[width=1.0\textwidth]{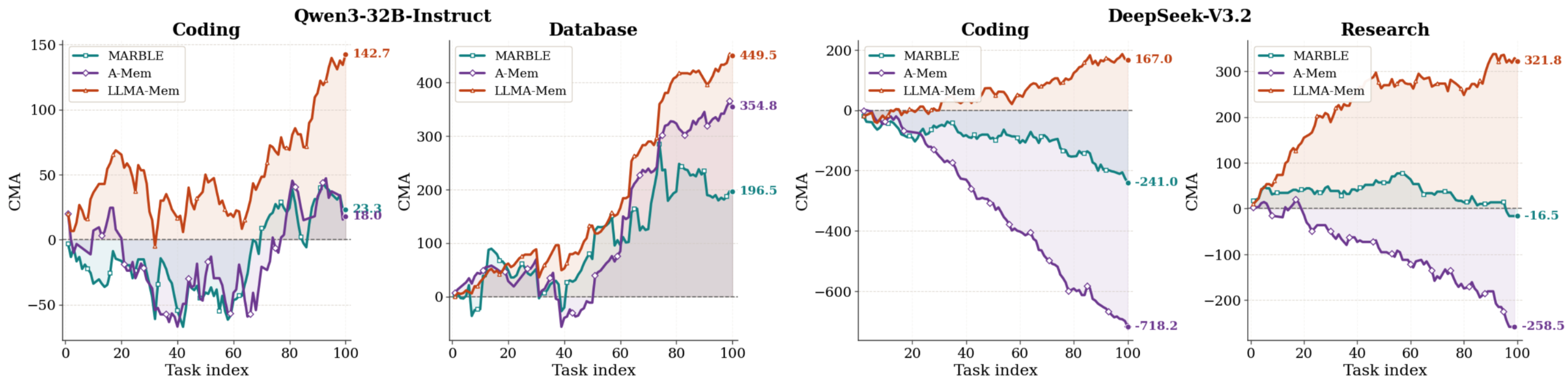}
\caption{Cumulative moving average (CMA) curves on representative settings. LLMA-Mem shows more stable long-horizon improvement than MARBLE and A-Mem, with especially margins represented by DeepSeek-v3.2 and Qwen3-32B-Instruct.}
\label{fig: CMA}
\end{figure}

From the lifelong learning perspective, \textit{AAS} shows a clearer advantage for LLMA-Mem. LLMA-Mem consistently improves over the no-memory baseline. The largest gains appear on DeepSeek-v3.2 in Research \gain{5.92} and Qwen3-32B-Instruct in Database \gain{3.19}, showing that the framework can convert accumulated experience into sustained performance improvements over long task sequences. The only exception is Qwen3-Next-80B on Research, where LLMA-Mem raises TS but reduces CS enough to slightly lower AAS, implying that better memory alone is not sufficient if coordination quality does not keep pace. Figure~\ref{fig: CMA} further illustrates this long-horizon behavior at the trajectory level. On Qwen3-32B-Instruct, LLMA-Mem finishes with the highest CMA in both Coding and Database compare to the baselines. The contrast is even sharper on DeepSeek-v3.2: LLMA-Mem remains positive throughout most of the sequence and ends at 167.0 on Coding and 321.8 on Research, while both MARBLE and A-Mem drift substantially negative by the end of the run. This indicates that LLMA-Mem improves average final performance, also yields a more stable accumulation of gains over time, whereas baseline memory designs are more vulnerable to performance decay as tasks progress.

\subsection{Cost Analysis}
Figure~\ref{fig: scaling space and cost comparison} shows that LLMA-Mem is not only effective in long-horizon performance, but also consistently more token-efficient than the compared memory baselines. Across all model--environment pairs, LLMA-Mem yields lower average token usage than MARBLE and A-Mem, with reductions ranging from \costdown{9.4\%} to \costdown{71.7\%}. The savings are especially large on Claude-Sonnet-4.5 for Research \costdown{60.4\%} and Coding \costdown{42.2\%}, on DeepSeek-V3.2 for Coding \costdown{47.5\%}, on Qwen3-32B-Instruct for Research \costdown{71.7\%} and Database \costdown{44.2\%}, and on Qwen3-next-80B for Coding \costdown{53.2\%}. Even in the smallest-margin cases, such as DeepSeek-V3.2 on Database \costdown{9.4\%} and Qwen3-next-80B on Database \costdown{14.5\%}, LLMA-Mem still maintains a clear cost advantage.

This reduction mainly comes from lower input-token overhead. In MARBLE and A-Mem, memory retrieval often injects larger context blocks into the prompt, causing token cost to grow substantially as more experiences accumulate. By contrast, LLMA-Mem consolidates episodic traces into compact procedural memories, allowing the system to reuse past experience in a more compressed form rather than repeatedly replaying raw trajectories. As a result, LLMA-Mem avoids much of the context inflation suffered by the other memory baselines while preserving useful long-term knowledge. Taken together with the main-results discussion, these findings highlight an important property of LLMA-Mem: its gains are not obtained by simply spending more tokens. Instead, LLMA-Mem achieves a better efficiency--effectiveness trade-off, improving performance over baselines while simultaneously reducing token consumption relative to alternative memory designs. This supports our central claim that better memory consolidation can scale time more effectively than naively scaling context.

\begin{table}[htbp!]
\centering
\small
\setlength{\tabcolsep}{4pt}
\renewcommand{\arraystretch}{1.15}
\caption{Influence of team size on performance. We report average task score (TS), coordination score (CS), Average of AS (AAS), and average token usage per task.}
\label{tab:team_size}
\begin{tabular}{lccccc}
\toprule
\textbf{Model} & \textbf{Team Size} & \textbf{Avg TS} & \textbf{Avg CS} & \textbf{AAS} & \textbf{Avg Tokens} \\
\midrule

\multirow{4}{*}{Claude-Sonnet-4.5}
& 1 & 77.08 & 49.58 & 57.78 & 53,549 \\
& 3 & 76.89 & 48.78 & 58.96 & 139,948 \\
& 5 & 77.33 & \cellcolor{lightgreen}\best{58.00} & \cellcolor{lightgreen}\best{62.57} & 279,563 \\
& 7 & \cellcolor{lightgreen}\best{78.22} & 53.11 & 62.21 & 344,323 \\

\midrule

\multirow{4}{*}{DeepSeek-V3.2}
& 1 & 72.08 & 48.13 & 49.01 & 41,205 \\
& 3 & 72.44 & 52.00 & 54.48 & 116,409 \\
& 5 & \cellcolor{lightgreen}\best{73.75} & 50.83 & 57.76 & 167,755 \\
& 7 & 73.33 & \cellcolor{lightgreen}\best{53.65} & \cellcolor{lightgreen}\best{58.79} & 274,360 \\

\midrule

\multirow{4}{*}{Qwen3-next-80B}
& 1 & 60.95 & 40.71 & 35.59 & 44,251 \\
& 3 & 57.95 & 35.13 & 34.47 & 127,120 \\
& 5 & 63.33 & 38.75 & 43.27 & 192,425 \\
& 7 & \cellcolor{lightgreen}\best{75.71} & \cellcolor{lightgreen}\best{41.43} & \cellcolor{lightgreen}\best{55.05} & 275,234 \\

\midrule

\multirow{4}{*}{Qwen3-32B-Instruct}
& 1 & 72.31 & 47.44 & 53.22 & 45,247 \\
& 3 & \cellcolor{lightgreen}\best{72.86} & 54.64 & 59.21 & 107,808 \\
& 5 & 70.00 & \cellcolor{lightgreen}\best{56.96} & 58.45 & 216,170 \\
& 7 & 71.67 & 54.79 & \cellcolor{lightgreen}\best{60.98} & 228,069 \\

\bottomrule
\end{tabular}
\end{table}

\subsection{Influence of team size}
Supported by the original \textsc{MultiAgentBench} settings, we conduct experiments with LLMA-Mem using team sizes of \{1, 3, 5, 7\} on a subset of 16 tasks from the research environment that support at least seven agents, in order to investigate the influence of team size. We keep the same graph communication topology and set the memory consolidation interval to 3 for LLMA-Mem to better observe evolving behaviors over time. Figure~\ref{fig: scaling space and cost comparison} illustrates how lifelong learning ability is affected by team size. From a temporal perspective, LLMA-Mem demonstrates consistent performance improvement as the task index increases across all team sizes and models, indicating that the system benefits from accumulated experience during sequential tasks. From a spatial perspective, increasing team size generally leads to improved performance as well. This two-dimensional perspective naturally defines \textbf{scaling space} of LLM multi-agent systems.

Interestingly, this scaling space is non-monotonic. For example, with Qwen3-32B-Instruct we observe a performance inversion where a 3-agent team outperforms a 5-agent team. This phenomenon is even more pronounced for Claude-Sonnet-4.5, where the 5-agent team surpasses the 7-agent team after approximately the fifth task. This observation leads to an important insight: \textbf{a MAS with fewer agents can achieve superior long-term performance through stronger lifelong learning ability}. Essentially, this reflects a trade-off between performance and computational cost in long-horizon scenarios. To further analyze this trade-off, we examine the relationship between average token cost per task and performance across different team sizes and models. As shown in Table \ref{tab:team_size}, increasing team size consistently results in higher computational cost across all models, and this cost growth becomes more pronounced as model scale increases. Meanwhile, the trends of TS and CS are not always aligned. For example, in both DeepSeek-V3.2 and Qwen3-32B-Instruct, TS achieves higher values at smaller team sizes than those corresponding to the best CS results. This suggests that communication overhead can become a limiting factor in MAS as team size increases.

Furthermore, when model scale is also considered as part of the cost, an interesting result emerges. Benefiting from the lifelong learning ability enabled by LLMA-Mem, Qwen3-32B-Instruct achieves a task score (TS) advantage at a 3-agent team despite having significantly fewer parameters than the other models. This configuration represents a local optimum in the cost–performance trade-off among all evaluated settings.

\begin{table}[htbp!]
\centering
\small
\setlength{\tabcolsep}{7pt}
\caption{Ablation results on the coding environment with Qwen3-32B-Instruct. The upper block compares memory topologies under a fixed consolidation interval $N{=}5$. The lower block compares consolidation intervals under the local topology.}
\label{tab:ablation_coding}
\begin{tabular}{llccc}
\toprule
\textbf{Study} & \textbf{Setting} & \textbf{TS} & \textbf{CS} & \textbf{AAS} \\
\midrule
\multirow{3}{*}{Topology ($N{=}5$)}
& Local  & \cellcolor{lightgreen}\best{49.68} & \cellcolor{lightgreen}\best{46.69} & \cellcolor{lightgreen}\best{49.43} \\
& Shared & 46.79 & 45.96 & 48.62 \\
& Hybrid & 46.52 & 40.32 & 46.92 \\
\midrule
\multirow{4}{*}{Consolidation (Local)}
& $N{=}2$  & 45.86 & 42.93 & 45.07 \\
& $N{=}5$  & \cellcolor{lightgreen}\best{49.68} & \cellcolor{lightgreen}\best{46.69} & \cellcolor{lightgreen}\best{49.43} \\
& $N{=}10$ & 47.72 & 42.36 & 46.87 \\
& $N{=}20$ & 47.71 & 41.43 & 47.30 \\
\bottomrule
\end{tabular}
\end{table}

\subsection{Ablation Study}

\subsubsection{Influence of Memory Topology}
Table~\ref{tab:ablation_coding} summarizes the ablation results. To isolate the effect of topology, we compare the three topologies under the same consolidation interval $N{=}5$. Under this controlled comparison, the local topology achieves the best performance on all three metrics, outperforming the shared and hybrid topology. The gap is especially clear on AAS, indicating that the local design yields the strongest long-horizon learning effect. We attribute this advantage to role-specific memory formation. In the coding environment, different agents often play different functional roles, such as decomposition, implementation, debugging, and verification. A local memory store allows each agent to accumulate procedures and experiences that are tightly coupled to its own role, which makes subsequent retrieval more compatible and more directly reusable. By contrast, in the shared topology, experiences generated under one role may later be retrieved by another agent operating under a different responsibility, leading to incompatible memory reuse and weaker downstream coordination. The hybrid topology does not fully eliminate this problem, because the shared component still introduces cross-role interference. Overall, these results suggest that, in the lifelong setting, preserving role-specific memories is more beneficial than maximizing global sharing, making the local topology the most effective memory topology among the tested designs.
\subsubsection{Influence of Memory Consolidation interval}
We next analyze the effect of the consolidation interval $N$ under the local topology, comparing $N \in \{2, 5, 10, 20\}$. As shown in Table~\ref{tab:ablation_coding}, the best overall setting is $N{=}5$, while both more frequent consolidation ($N{=}2$) and less frequent consolidation ($N{=}10, 20$) perform worse. This pattern indicates that the consolidation interval exhibits a non-monotonic effect, with a moderate update frequency providing the best balance. Our interpretation is that $N{=}5$ offers a suitable granularity for procedural abstraction. When consolidation is too frequent, as in $N{=}2$, episodic experiences may be converted into procedures before enough evidence has accumulated, resulting in unstable or overly specific procedural memories. When consolidation is too sparse, as in $N{=}10$ or $20$, useful experiences remain trapped in episodic form for too long, delaying their reuse and weakening the cumulative benefit over the task sequence. In contrast, $N{=}5$ appears to provide enough repeated evidence to form reliable procedures while still updating memory quickly enough to support future tasks. Therefore, under the local topology, a moderate consolidation schedule is the most effective design choice, and together with the topology result, the ablation study suggests that local memory topology plus moderately frequent consolidation is the strongest configuration for lifelong learning scenarios.

\section{Conclusion}
In this paper, we study LLM multi-agent systems from a joint scaling perspective that considers both scaling across agents and scaling across time. Rather than treating team size and lifelong learning as separate questions, we argue that they form an interacting scaling space in which long-horizon utility depends on how effectively a system converts experience into reusable knowledge under coordination and cost constraints. From this perspective, memory is not a peripheral add-on; it is the mechanism that reshapes how multi-agent systems scale. To realize this idea, we propose \textbf{LLMA-Mem}, a lifelong memory framework that integrates episodic, procedural, and transactive memory under flexible memory topologies. This decomposition clarifies which aspects of our design primarily support lifelong learning and which are specifically required for long-horizon multi-agent collaboration.

Empirically, across coding, research, and database environments on \textsc{MultiAgentBench}, LLMA-Mem consistently improves long-horizon performance while reducing token cost by \costdown{9.4\%} to \costdown{71.7\%} relative to competing memory baselines. More importantly, our analysis reveals that the interaction between team size and lifelong learning is fundamentally non-monotonic. Larger teams can provide stronger parallelism, but they also introduce communication overhead and information fragmentation that can weaken long-term gains. As a result, smaller teams can outperform larger ones when memory better supports the stable accumulation and reuse of experience.

\section{Limitations}
Our work still has several limitations. First, although our experiments demonstrate clear lifelong learning capabilities and reveal a synergistic interaction with team size, the explored team scale remains limited. While we analyze multiple model families and vary team size up to seven agents, this range may not fully capture the behavior of larger collectives, where communication bottlenecks, role specialization, and memory interference are likely to become more pronounced. Second, our evaluation is restricted to three collaborative environments in \textsc{MultiAgentBench}: coding, research, and database. Although these settings are representative for studying sequential agent interaction, they do not encompass other important lifelong scenarios such as web search, embodied control, or personalization. Finally, our study focuses on overall task and communication performance, but does not explicitly assess memory quality itself, such as redundancy, staleness, or robustness to retrieval errors. We consider these directions important avenues for future work toward a more comprehensive understanding of lifelong learning in multi-agent systems.

\bibliographystyle{abbrvnat}
\nobibliography*
\bibliography{custom}

\appendix

\section{Details of LLMA-Mem Design}
\label{app:llmamem_design}
Our LLMA-Mem implementation follows the three-part design described in Section \ref{sec: method}, consisting of episodic, procedural, and transactive memory. In the main experiments, the default retrieval setting uses the top-$3$ memory items as memory context for downstream generation. Unless otherwise specified, the main comparison in Table~\ref{tab:main results} uses the local topology with consolidation interval $N{=}5$, which is also the best-performing setting in the ablation study. For the team-size experiment in Section \ref{sec: experiments}, we set the consolidation interval to $N{=}3$ to expose memory evolution more frequently over the shorter task sequence. For the three topologies, LLMA-Mem is instantiated as follows. In the \emph{local} topology, each agent maintains a private memory store and writes only to its own episodic, procedural, and transactive memories. In the \emph{shared} topology, all agents read from and write to a common memory store. In the \emph{hybrid} topology, episodic memory remains private to each agent, while procedural memory and shared coordination statistics are globally accessible. Memory persistence is enabled through task-specific storage directories which store all memory into JSON files so that memories accumulate across tasks within the same sequential run. The memory lifecycle also follows implementation rules. After each task, a new episodic record is appended, used procedures update their success/failure counts, and transactive statistics are refreshed from the latest execution trace. Every $N$ tasks, episodic experiences are consolidated into candidate procedures by clustering semantically similar lessons learned, extracting generalized successful strategies, and removing redundant procedures whose supporting episodes are strict subsets of stronger procedures. This implementation corresponds directly to Algorithm~\ref{alg:procedure_extraction} in the main paper. This algorithm is designed as an abstract framework that can be applied across a wide range of environments. During the implementation on \textsc{MultiAgentBench}, the coding and research environments provide evaluation metrics in terms of TS and CS, while the database environment returns intermediate success signals. To ensure consistent adaptation of the algorithm, we define a threshold of 0.60 based on $S_t$ for the coding and research environments.

\begin{figure}[htbp!]
\centering
\includegraphics[width=1.0\textwidth]{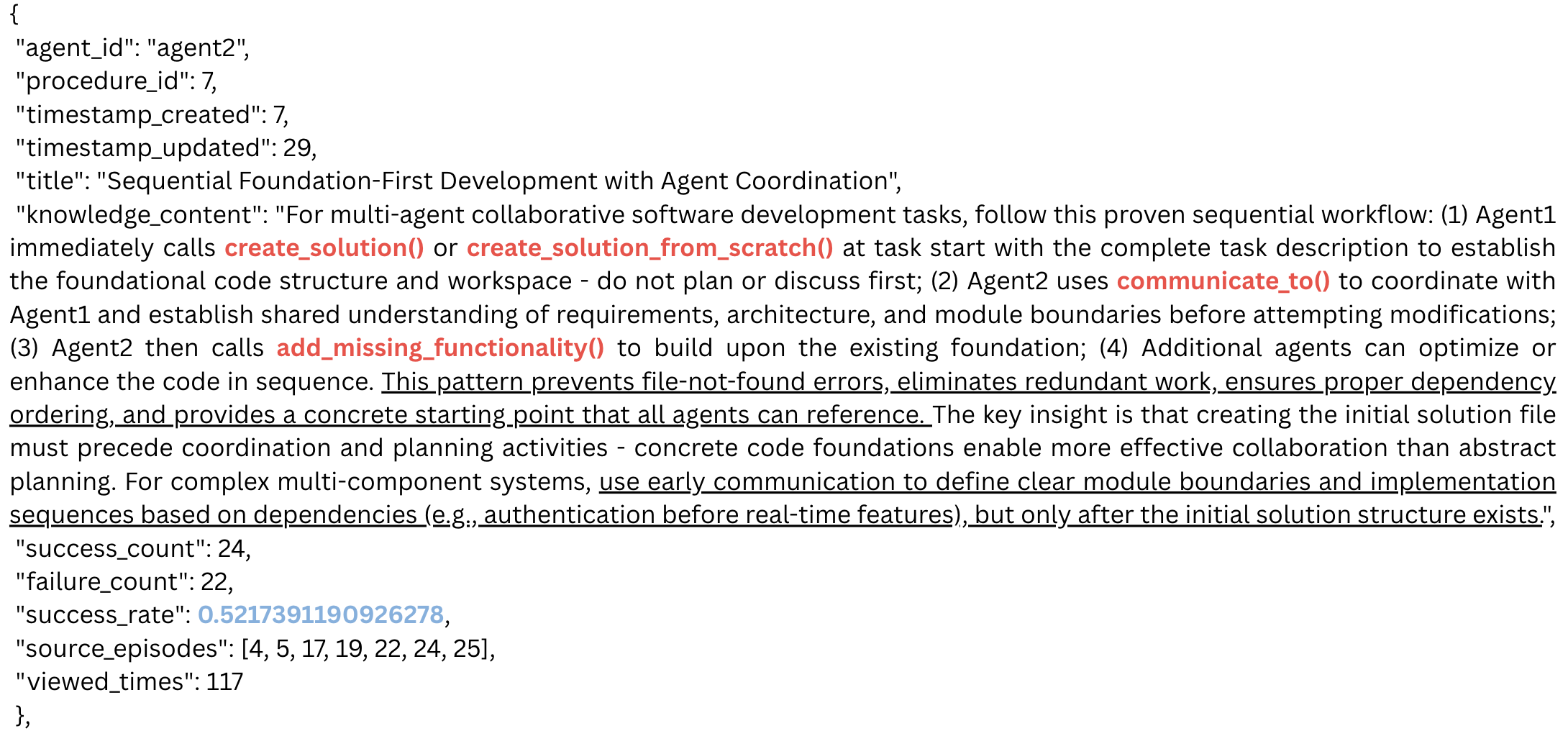}
\caption{An example of procedural memory piece from Qwen3-32B-Instruct on coding envrionment.}
\label{fig: example}
\end{figure}

Figure \ref{fig: example} shows an example of procedural memory piece from Qwen3-32B-instruct on coding envrionment for agent-2 in local memory topology. This example clearly illustrates what procedural memory looks like in practice within LLMA-Mem: it is a compact, structured, and reusable strategy distilled from multiple past task trajectories, rather than raw experience. Instead of storing full interaction logs, the memory encodes a generalized workflow pattern, such as "foundation-first development with coordination", along with success/failure counts, timestamps, and source episodes. As a knowlewdge content, LLMA-Mem explicitly guides tool-use and gives advice on workflow construction. It also illustrates the specific categories of failure that its strategy is designed to preempt. As a result, agents no longer need to re-derive coordination strategies from scratch or retrieve long trajectories, but can directly apply a distilled, high-level procedure. Moreover, because this procedure is grounded in multiple successful episodes, it represents validated behavioral priors rather than heuristic guesses.

\section{Details of \textsc{MultiAgentBench} Implementation}
\label{app:mab_details}
We use the collaborative environments of \textsc{MultiAgentBench} with graph-based coordination throughout all experiments. The three environments considered in this paper are coding, research, and database. We set the maximum iteration as three for all environments. For the team-size study, we select 16 research tasks that support at least seven agents, enabling a controlled comparison across team sizes $\{1,3,5,7\}$. All task instances are executed under the same evaluation pipeline provided by the benchmark. Following the original benchmark protocol, task outcomes are scored by the evaluation model, and we report both task score (TS) and communication score (CS). For research tasks, the benchmark evaluator rates innovation, safety, and feasibility, which are then normalized by the benchmark code into the reported task score. In our experiments, Claude-Sonnet-4.5 serves as the unified evaluator across all model backbones to reduce scoring variance introduced by different judges. Across all experiments, the agent communication structure is the graph topology specified in the released benchmark configurations. Model execution is conducted through Amazon Bedrock. We set temprature=0.7 and max\_seq\_length=1024 for LLM generation. For coding generations, we set temperature=0.1 and max\_seq\_length=4096. For LLMA-Mem and A-Mem, we use Titan-text-embeddings-v2 for embedding-based retrieval. We keep benchmark task content, agent profiles, and environment rules unchanged, and only vary the memory mechanism or team size depending on the experiment.

\section{Experiments Implementation Details}
\label{app:baseline_details}
\textbf{W.o. Memory.} This baseline disables memory usage entirely and uses the same benchmark task definitions, agent profiles, and graph communication structure. It therefore isolates the contribution of memory by keeping the multi-agent workflow unchanged.

\textbf{MARBLE.} We use MARBLE, the original \textsc{MultiAgentBench} memory-enabled setup as the baseline. In this setting, agents interact through the benchmark's default shared-memory mechanism. This baseline is intended to represent a classical memory-equipped multi-agent system with conventional memory support.

\textbf{A-Mem.} We implement A-Mem using the official repository's A-Mem configuration with Titan-text-embeddings-v2 as the embedding model, retrieval top-$3$, maximum memory context of $5$, and link threshold $0.72$. This baseline maintains an evolving associative memory structure and is therefore more comparable to LLMA-Mem. For fairness, A-Mem uses the same underlying agent tasks, models, evaluator, and benchmark environments as the LLMA-Mem runs.

\section{Prompt and Evaluation Notes}
\label{app:prompts}
This appendix collects the prompts used by LLMA-Mem. We distinguish between: (1) inherited benchmark task prompt templates from \textsc{MultiAgentBench}, which vary by task instance and are not modified by us, and (2) LLMA-Mem-specific prompt templates that govern memory retrieval, lesson extraction, and procedural consolidation. To improve readability, we show the implementation-level templates with placeholders.

\subsection{Inherited Benchmark Task Prompt}
The base task prompt for coding, research, and database experiments is inherited directly from the released \textsc{MultiAgentBench} task configurations. Each benchmark task provides the task objective, environment specification, output format, and agent profiles. LLMA-Mem does not rewrite these benchmark prompts; instead, it prepends memory context and uses the same downstream task description as the original benchmark.

\begin{tcolorbox}[breakable, colback=blue!2, colframe=emory, title={Task Prompt Structure}]
\small
\texttt{
task\_content := benchmark-provided task description \\
agent\_profile := benchmark-provided role/profile text \\
environment\_rules := benchmark-provided environment specification \\
output\_format := benchmark-provided output requirements
}
\end{tcolorbox}

\subsection{Agent Action Prompt}
At action time, LLMA-Mem constructs a single user prompt that combines the agent profile, optional reasoning strategy prompt, retrieved memory context, the current task, and the list of other agents. The following template is the core action prompt used by the agent:

\begin{tcolorbox}[breakable, colback=teal!2, colframe=teal!60!black, title={LLMA-Mem Agent Action Prompt}]
\begin{lstlisting}[basicstyle=\ttfamily\small, breaklines=true]
You are {agent_id}: {agent_profile}
{reasoning_prompt}
--- Past Experience ---
{memory_str}
--- End Past Experience ---

=== CURRENT TASK ===
{task}
=== END TASK ===

Other agents you can interact with:
{agent_descriptions}
You do not have to communicate with other agents.
\end{lstlisting}
\end{tcolorbox}

\subsection{Inserted Memory Context}
The retrieved memory block inserted into the action prompt differs by memory type. For LLMA-Mem, episodic retrieval and procedural retrieval are formatted as follows.

\begin{tcolorbox}[breakable, colback=green!2, colframe=green!50!black, title={Episodic Memory Prompt Block}]
\begin{lstlisting}[basicstyle=\ttfamily\small, breaklines=true]
[Past Experiences]
- Similar past task: {task_description}
  Result: {succeeded_or_had_issues}
  Takeaway: {lesson_1; lesson_2; ...}
\end{lstlisting}
\end{tcolorbox}

\begin{tcolorbox}[breakable, colback=green!2, colframe=green!50!black, title={Procedural Memory Prompt Block}]
\begin{lstlisting}[basicstyle=\ttfamily\small, breaklines=true]
[Relevant Procedures & Strategies]
- {procedure_title} (success rate: {success_rate})
  Strategy : {knowledge_content}
\end{lstlisting}
\end{tcolorbox}

\subsection{Lesson Extraction Prompt}
After each task, LLMA-Mem converts the execution trace into compact lessons for episodic memory. This prompt is used to extract one to three actionable lessons from the task outcome.

\begin{tcolorbox}[breakable, colback=purple!2, colframe=darklav, title={Lesson Extraction System Prompt}]
\begin{lstlisting}[basicstyle=\ttfamily\small, breaklines=true]
You are a reflective learning assistant that extracts actionable lessons from task experiences.
\end{lstlisting}
\end{tcolorbox}

\begin{tcolorbox}[breakable, colback=purple!2, colframe=darklav, title={Lesson Extraction User Prompt}]
\begin{lstlisting}[basicstyle=\ttfamily\small, breaklines=true]
Based on the following task experience, extract 1-3 concise, actionable lessons learned.
Focus on CONCRETE actions: which tool functions should have been called, what patterns worked or failed, and what specific steps to take next time.
Do NOT suggest vague advice like 'communicate better' or 'provide clearer instructions'.
Instead, suggest specific tool calls or strategies.

{optional_database_role_section}
Task: {task_description}

{optional_task_summary_context}
Actions taken: {actions_str}
Outcome: {outcome_str}

{success_or_failure_focus}
Return the lessons as a JSON array of strings. Example:
["Lesson 1", ...]
\end{lstlisting}
\end{tcolorbox}

\subsection{Procedural Consolidation Prompt}
Every $N$ tasks, LLMA-Mem consolidates successful episodic experiences into reusable procedures. The following prompt is used to generalize a strategy from successful episodes.

\begin{tcolorbox}[breakable, colback=red!2, colframe=red!60!black, title={Procedure Generalization System Prompt}]
\begin{lstlisting}[basicstyle=\ttfamily\small, breaklines=true]
You extract generalized, reusable strategies from task experiences.
\end{lstlisting}
\end{tcolorbox}

\begin{tcolorbox}[breakable, colback=red!2, colframe=red!60!black, title={Procedure Generalization User Prompt}]
\begin{lstlisting}[basicstyle=\ttfamily\small, breaklines=true]
Based on the following successful task experiences, extract a generalized
and actionable strategy and skill that can be reused in similar future situations.
Avoid vague advice.

Episode 1:
  Task: {task_description_1}
  Lessons: {lessons_1}
  Outcome: {outcome_1}

Episode 2:
  Task: {task_description_2}
  Lessons: {lessons_2}
  Outcome: {outcome_2}

...

Respond in JSON format:
{"title": "Short descriptive title",
 "knowledge_content": "Detailed strategy and skill description"}
\end{lstlisting}
\end{tcolorbox}

\end{document}